\documentclass[journal]{IEEEtran}

\usepackage{graphicx,epsfig}
\usepackage{amsmath,mathrsfs}
\usepackage{algorithm}
\usepackage{algpseudocode}
\makeatletter
\def\BState{\State\hskip-\ALG@thistlm}
\makeatother
\usepackage{amssymb}
\usepackage{amsthm}
\usepackage{amsfonts}
\usepackage{mathtools}
\usepackage{multirow}
\usepackage{mathtools} 
\usepackage{cite}
\usepackage{epstopdf}
\usepackage{ifthen}
\usepackage{bbm}
\usepackage{setspace}

\newcommand{\R}{ \mathbb{R}}

\newcommand{\N}{ \mathbb{N}}

\newcommand{\ud}{\,\mathrm{d}}

\def\V#1{{\boldsymbol{#1}}}         
\def\M#1{{\bf{#1}}}  

\newtheorem{theorem}{Theorem}

\newtheorem{definition}[theorem]{Definition}

\makeatletter
\def\blfootnote{\xdef\@thefnmark{}\@footnotetext}
\makeatother

\usepackage{tikz,pgfplots}
\pgfplotsset{compat=newest}
\usetikzlibrary{intersections,calc,arrows,matrix,spy,math,calc,fit,positioning, decorations.pathmorphing, tikzmark}
\tikzset{snake it/.style={decorate, decoration=snake}}
\pgfplotsset{compat=newest,legend style={font=\footnotesize},
                     ticklabel style={font=\footnotesize},
                     x label style={font=\footnotesize},
                     y label style={font=\footnotesize}}

\usetikzlibrary{arrows.meta,shapes.geometric}

\usepackage{ dsfont,color }
\usepackage[nottoc, notlof, notlot]{tocbibind}
\usepackage{bbm}
\usepackage{stackengine}
\usepackage{graphicx}
\usepackage[font=normalsize]{subfig}
\usepackage{booktabs}

\usepackage[final]{changes}
\makeatletter
\@namedef{Changes@AuthorColor}{red}
\colorlet{Changes@Color}{red}
\makeatother

\usepackage{tikz}

\begin{document}

\IEEEoverridecommandlockouts

\newcommand{\rev}[1]{\textcolor{red}{#1}}

\title{A Statistical Framework to Investigate the Optimality of \added{Signal-Reconstruction Methods}}

\author{Pakshal~Bohra, Pol~{del Aguila Pla},~\IEEEmembership{Member,~IEEE}, Jean-Fran\c{c}ois~Giovannelli, and~Michael~Unser,~\IEEEmembership{Fellow,~IEEE}
\thanks{This work was supported in part by the Swiss National Science Foundation under Grant 200020\_184646 / 1 and in part by the European Research Council (ERC Project FunLearn) under Grant 101020573.

P.~{del Aguila Pla} is with the CIBM Center for Biomedical Imaging, Switzerland.

P. Bohra, P.~{del Aguila Pla} and M. Unser are with the Biomedical Imaging Group, \'{E}cole polytechnique f\'{e}d\'{e}rale de Lausanne, 1015 Lausanne, Switzerland (e-mail: pakshal.bohra@epfl.ch; pol.delaguilapla@epfl.ch; michael.unser@epfl.ch).

J. F. Giovannelli is with IMS (Univ. Bordeaux, CNRS, B-INP), UMR 5218, F-33400 Talence, France (e-mail: jean-francois.giovannelli@u-bordeaux.fr).
}

}



\maketitle

\begin{abstract}
We present a statistical framework to benchmark the performance of reconstruction algorithms for linear inverse problems, in particular, neural-network-based methods that require large quantities of training data. We generate synthetic signals as realizations of sparse stochastic processes, which makes them ideally matched to variational sparsity-promoting techniques. We derive Gibbs sampling schemes to compute the minimum mean-square error estimators for processes with Laplace, Student's t, and Bernoulli-Laplace innovations. These allow our framework to provide quantitative measures of the degree of optimality (in the mean-square-error sense) for any given reconstruction method. We showcase our framework by benchmarking the performance of some well-known variational methods and convolutional neural network architectures that perform direct nonlinear reconstructions in the context of deconvolution and Fourier sampling. Our experimental results support the understanding that, while these neural networks outperform the variational methods and achieve near-optimal results in many settings, their performance deteriorates severely for signals associated with heavy-tailed distributions. 
\end{abstract}

\section{Introduction}\label{sec:intro}
 Inverse problems are often encountered in biomedical imaging \cite{unser2019biomedical}, particularly in modalities such as computed tomography (CT), magnetic resonance imaging (MRI), or deconvolution microscopy. Their goal is to reconstruct an unknown signal from its measurements. Often, these are hard to solve due to their ill-posedness, which implies that the underlying signal cannot be determined uniquely by the acquired measurements, unless one introduces some form of regularization. Therefore, prior knowledge about the signal of interest is required for the resolution of such problems.

\subsection{Model-Based Methods}
Model-based methods rely on the mathematical modeling of the signal of interest to counteract the ill-posedness of the inverse problem. We organize them in two categories. 

The first category is composed of linear reconstruction methods (\textit{e.g.,} filtered back-projection), which are fast, well understood, and come with performance and stability guarantees \cite{tikhonov1963, bertero1998}. From a variational standpoint, they can be interpreted as minimizers of a cost functional that consists of a quadratic data term to ensure consistency with the measurements, along with an additive quadratic (Tikhonov) regularization term that imposes some smoothness on the solution. Interestingly, these methods can also be derived from a statistical perspective as optimal linear reconstructors under the Gaussian hypothesis \cite{kay1993fundamentals}. 

The second category is composed of methods that exploit sparsity---the property that a signal admits a concise representation in some transform domain (\textit{e.g.,} wavelets) \cite{mallat1999wavelet, bruckstein2009sparse,baraniuk2010applications,elad2010sparse}. This powerful concept supports the theory of compressed sensing, which gives conditions under which the reconstruction of an image from a limited set of measurements is feasible \cite{donoho2006compressed,candes2008introduction,foucart2013invitation} and stable \cite{candes2006stable, pol2023stable}. To obtain a sparse reconstruction, one typically uses $\ell_1$-norm regularization and solves the corresponding convex optimization problem using iterative algorithms such as the fast iterative shrinkage-thresholding algorithm (FISTA) \cite{beck2009fast} or the alternating direction method of multipliers (ADMM) \cite{boyd2011distributed}. In practice, sparsity-promoting regularizers such as total variation (TV)~\cite{rudin1992nonlinear} generally improve the quality of the image. From a statistical point of view, many of these sparsity-based methods can be interpreted as maximum a posteriori (MAP) estimators for some specific choices of stochastic models for the signal of interest \cite{babacan2009bayesian}.

\subsection{Learning-Based Methods}
Neural-network-based methods that make use of prior information learned from a large collection of training data are now the focus of much of the current research in image reconstruction \cite{mccann2017convolutional,ongie2020deep}. They shine in extreme imaging scenarios where one wishes to achieve more with fewer data, for instance when operating with short integration times, which leads to an abundance of noise, or when collecting fewer measurements to reduce either the acquisition duration and/or the radiation exposure \cite{jin2017deep}. Here, we focus on two classes of neural-network-based methods and classify them as the counterparts of the model-based ones.

The first successful applications of deep convolutional neural networks (CNNs) in imaging build upon the classical linear-reconstruction algorithms, training a CNN to correct for reconstruction artifacts in extreme imaging conditions \cite{jin2017deep,chen2017low,hyun2018deep,monakhova2019learned,perdios2021cnn}. Unrolling methods \cite{gregor2010learning,chen2016trainable,sun2016deep,aggarwal2018modl,adler2018learned,monga2021algorithm} also fall into this class of direct, nonlinear reconstructions. Examples of successful applications include MRI, CT, optical imaging, and ultrasound. Their gain over the state-of-the-art is impressive and comparable in magnitude to the one afforded by a decade of refinement of the sparsity-promoting techniques.

The second class includes methods that attempt to reconstruct an image that is consistent with the measurements by replacing the proximal operator that is typically involved in the iterative sparsity-promoting methods by an appropriate denoising CNN, which then plays the role of the regularizer. They come in a variety of flavors, including plug-and-play (PnP) \cite{venkatakrishnan2013plug,ryu2019plug,zhang2021plug,sun2021scalable}, regularization-by-denoising (RED) \cite{romano2017little,sun2019block,wu2020simba}, and projected-gradient-descent \cite{rick2017one,gupta2018cnn} methods.

Despite their remarkable performance, CNN-based imaging methods have limitations that currently hinder their further development. Unlike the model-based methods, which are backed by sound mathematics, the development of CNN-based approaches is empirical. Expressivity is obtained through the composition of simple units, but the working of the whole is hard to comprehend and the architectural options are overwhelming (\textit{e.g.,} depth, number of channels, size of the filters). In practice, one usually proceeds by trial and error using the training, validation, and testing errors as quantitative criteria. Further, the training of CNNs is poorly understood and often difficult because of the underlying over-parameterization: getting a stochastic optimization algorithm to perform properly for a specific application typically requires a lot of adjustments and experimentation. 

Beside the strain that this empirical approach exerts on developers, the performance greatly depends on the quality, cardinality, and representability of the
training dataset, while the outcome is not necessarily transposable to other applications. The bottleneck with biomedical imaging is often a limited access to large, representative datasets. This is mostly because of legal issues in medical imaging and because of the lack of standardized protocols in biomicroscopy. Another issue is the chicken-and-egg nature of the training process because the desired image (the physical object that corresponds to the measurements) is not known precisely---in practice, the goldstandard is an image produced by a state-of-the-art model-based method with high-density/low-noise measurements. This is adequate for developing methods for compressed sensing, but not otherwise. This explains why the works that demonstrate the superiority of the
CNN-based approaches over the more traditional model-based methods for image reconstruction have used limited benchmarks so far.

\subsection{Contribution}
In this work, we present an objective environment to benchmark the performance of reconstruction algorithms for linear inverse problems. Our proposed framework offers quantitative measures of the degree of optimality (in the mean-square-error sense) for any given reconstruction method. Further, it provides access to large amounts of training data, which enables the benchmarking of CNN-based approaches.

We synthesize ground-truth signals and then simulate the measurement process (\textit{e.g.,} convolution for deconvolution microscopy, Fourier sampling for MRI) in the presence of noise. Specifically, we consider a statistical framework where the underlying signals are realizations of 1D sparse stochastic processes (SSPs) \cite{unser2014ssp}. The motivation there is that these processes are ideally matched to model-based methods, the most prominent of which can be interpreted as their MAP estimators \cite{kamilov2012mmse}. Since the true statistical distribution of the signal is known exactly in our framework, the minimum-mean-square-error (MMSE) estimator is indeed optimal in the mean-square-error (MSE) sense. Therefore, we are able to provide statistical guarantees of optimality by specifying an upper limit on the reconstruction performance.

Our framework also provides training data for CNN-based approaches. Indeed, we can produce any desired number of training pairs for a given reconstruction task and some chosen stochastic signal model, which allows for an informed comparison of network architectures. Thus, the availability of the goldstandard (MMSE estimator) and training data make our benchmark a good ground for the tuning of CNN architectures and for the identification of the best designs in a tightly controlled environment.

The MAP estimates of SSPs are solutions of optimization problems that resemble the ones used in model-based methods, and can be computed efficiently. However, it has been observed that these MAP estimators are suboptimal in the MSE sense \cite{kamilov2012mmse,amini2012bayesian}, except in the Gaussian scenario where the MAP and MMSE estimators (generalized Wiener filter) coincide \cite{kay1993fundamentals}. In this work, we focus on non-Gaussian signal models. In principle, the MMSE estimator involves the calculation of high-dimensional integrals, which are not numerically tractable in general. Thus, we develop efficient Gibbs-sampling-based algorithms to compute the MMSE estimator for specific classes of SSPs, with innovations following the Laplace, Student's t, and Bernoulli-Laplace distributions. To the best of our knowledge, no such working solution for generic linear inverse problems with SSPs has been presented in the literature.

Finally, we present experimental results that illustrate the usefulness of our framework. Specifically, we benchmark the performance of some well-known model-based methods and CNNs that perform direct nonlinear reconstructions, in the context of deconvolution and Fourier sampling for first-order SSPs. The CNNs that we consider are optimized by minimizing the MSE loss for training datasets. On one hand, when the innovations follow a Bernoulli-Laplace distribution, we observe that CNNs (with sufficient capacity and training data) outperform the sparsity-promoting methods, which are well-suited to these piecewise-constant signals. In fact, some of these CNNs achieve near-optimal MSE performance. On the other hand, our experiments with Student's t innovations indicate regimes where CNNs fail to reconstruct the signals well. More specifically, we observe that, when the tails of the Student's t distribution are made heavier (\textit{i.e.,} when we move towards a Cauchy distribution), CNNs perform rather poorly.

\subsection{Roadmap}
In Section \ref{sec:measurement_model}, we describe a continuous-domain model for the measurement process along with a way to discretize it. In Section \ref{sec:stochastic_signal_model}, we introduce L\'{e}vy processes as stochastic models for our signals and we derive the probability distribution for samples of such processes. We then discuss MAP and MMSE estimation in Section \ref{sec:bayesian_inference} before we develop Gibbs samplers for L\'{e}vy processes associated with Laplace, Student's t, and Bernoulli-Laplace distributions in Section \ref{sec:mmse_gibbs}. Finally, we present experimental results in Section \ref{sec:experiments}.

\section{Measurement Model}\label{sec:measurement_model}
In the proposed framework, we consider the recovery of a continuous-domain signal $s:\R \rightarrow \R$ from a finite number $M$ of measurements $\M y = (y_m)_{m=1}^{M}$.

\subsection{Continuous-Domain Measurement Model}
We model the measurements $\M y = (y_m)_{m=1}^{M}$ as
\begin{align}\label{eq:cont_dom_prob}
    y_m &=  \int_{\R} s(t)\nu_m(t) \ud t + n[m],
\end{align}
where $(\nu_m)_{m=1}^{M}$ are linear functionals that describe the physics of the acquisition process and $n[\cdot]$ is an additive white Gaussian noise (AWGN) with variance $\sigma_{\mathrm{n}}^{2}$. By choosing appropriate functionals $(\nu_m)_{m=1}^{M}$, we can study a variety of linear inverse problems such as denoising, deconvolution, inpainting, and Fourier sampling.

\subsection{Discrete Measurement Model}
We need to discretize \eqref{eq:cont_dom_prob} to obtain a computationally feasible model for the measurements. To that end, we consider a finite region of interest $\Omega = (0,T)$ of the signal and approximate it with
\begin{equation}\label{eq:fin_rep}
    s_h(t) = \sum\limits_{k=1}^{K} s(kh)\text{sinc}\Big(\frac{t}{h} - k\Big),
\end{equation}
where $h$ is the sampling step and $K = \big(\left\lfloor \frac{T}{h} \right \rfloor - 1 \big)$. When $h$ is small enough, $s_h$ is a good approximation of $s$ within the interval $\Omega$ \cite{unser2000sampling}. On introducing \eqref{eq:fin_rep} into \eqref{eq:cont_dom_prob}, we get that
\begin{equation}\label{eq:discrete_1}
    \M y = \M H \M s + \M n,
\end{equation}
where $\M s = (s(kh))_{k=1}^{K} \in \R^K$ contains equidistant samples of the signal, $\M H: \R^K \rightarrow \R^M$ is the discrete system matrix with
\begin{equation}
    [\M H]_{m,k} = \int_{\R} \text{sinc}\Big(\frac{t}{h} - k\Big) \nu_m(t) \ud t,
\end{equation}
and $\M n \in \R^M$ is the noise.

Thus, for any signal samples $\M s \in \R^{K}$, we can simulate noisy measurements using \eqref{eq:discrete_1}. Next, we derive the discrete system matrices for deconvolution and Fourier sampling. Hereafter, we assume for simplicity that $h=1$. 

\subsection{Deconvolution}\label{sec:deconv_forward}
In deconvolution, the measurements are acquired by sampling the result of the convolution between the signal and the point-spread function (PSF) $\psi$ of the acquisition system, which we model by letting the measurement functionals be $\nu_m = \psi(m - \cdot)$. We assume that the cutoff frequency of $\psi$ is $\omega_0 \leq \pi$, as this allows us to sample $(s*\psi)$ on an integer grid without aliasing effects. In this case,  The entries of the resulting system matrix $\M H$ are given by
\begin{align}
    [\M H]_{m,k} &= \int_{\R} \text{sinc}(t - k) \psi(m - t) \ud t \nonumber \\
                 &= \psi(m-k).
\end{align} 
Here, $\M H$ is a discrete convolution matrix whose entries are samples of the bandlimited PSF $\psi$.

\subsection{Fourier Sampling}\label{sec:fourier_sampling_forward}
In Fourier sampling, the measurements are acquired by sampling the Fourier transform of the signal at arbitrary frequencies $\{\omega_m\}_{m=1}^{M}$. Accordingly, the measurement functionals are the complex exponentials $\nu_m = \mathrm{e}^{-\mathrm{j} \omega_m \cdot}$. Assuming that $|\omega_m| \leq \pi$, we get that
\begin{align}
    [\M H]_{m,k} &= \int_{\R} \text{sinc}(t - k) \mathrm{e}^{-\mathrm{j} \omega_m t} \ud t \nonumber \\
                 &= \mathrm{e}^{-\mathrm{j} \omega_m k}.
\end{align} 
Here, $\M H$ is a discrete Fourier-like matrix, except that the frequencies $\omega_m$ do not necessarily lie on an uniform grid.

\section{Stochastic Signal Model}\label{sec:stochastic_signal_model}
In this section, we describe a continuous-domain stochastic model for the signal. We also derive the probability distribution for the discrete signal vector $\M s = (s(k))_{k=1}^{K}$.

\subsection{L\'{e}vy Processes}
In our framework, the underlying signals are realizations of a well-known class of first-order sparse stochastic processes: the L\'{e}vy processes \cite{ken1999levy,unser2014ssp}. 

\begin{definition}[L\'{e}vy process]
A stochastic process $s = \{s(t) : t \in \R^{+}\}$ is a L\'{e}vy process if
\begin{enumerate}
    \item $s(0) = 0$ almost surely;
    \item (independent increments) for any $N \in \mathbb{N}\setminus\{0,1\}$ and $0 \leq t_1 < t_2 \cdots < t_N < \infty$, the increments $\big(s(t_2)-s(t_1)\big), \big(s(t_3)-s(t_2)\big), \ldots, \big(s(t_N)-s(t_{N-1})\big)$ are mutually independent;
    \item (stationary increments) for any given step $h$, the increment process $u_h = \{s(t) - s(t-h) : t \in \R^{+}\}$ is stationary;
    \item (stochastic continuity) for any $\epsilon > 0$ and $t \geq 0$
    \begin{equation*}
        \lim_{h \rightarrow 0} \textup{Pr}\{|s(t+h)-s(t)| > \epsilon\} = 0.
    \end{equation*}
\end{enumerate}
\end{definition}

\begin{figure}[t]
    \subfloat[Gaussian: $p(x) = \frac{1}{\sqrt{2\pi\sigma^2}}\mathrm{e}^{-\frac{x^2 \ }{2\sigma^2}}$]{\includegraphics[trim={2.5cm 0 2cm 0}, clip,width=\columnwidth]{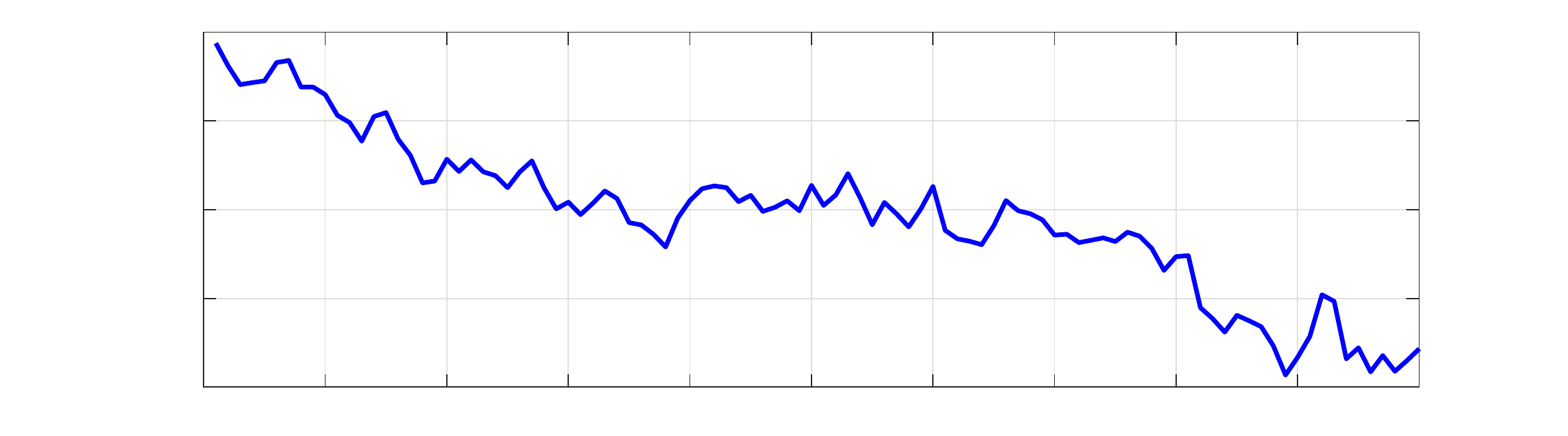}}

    \subfloat[Laplace: $p(x) = \frac{b}{2}\mathrm{e}^{-b |x|}$]{\includegraphics[trim={2.5cm 0 2cm 0}, clip,width=\columnwidth]{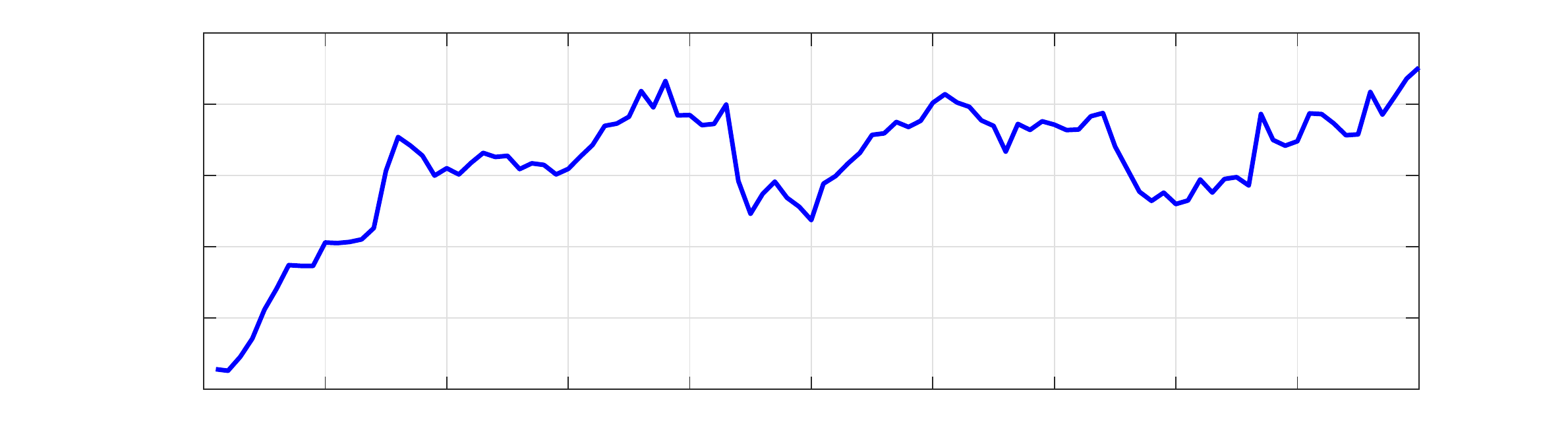}}

    \subfloat[Bernoulli-Laplace: $p(x) = \lambda \delta(x) + (1-\lambda)\frac{b}{2}\mathrm{e}^{-b |x|}$]{\includegraphics[trim={2.5cm 0 2cm 0}, clip,width=\columnwidth]{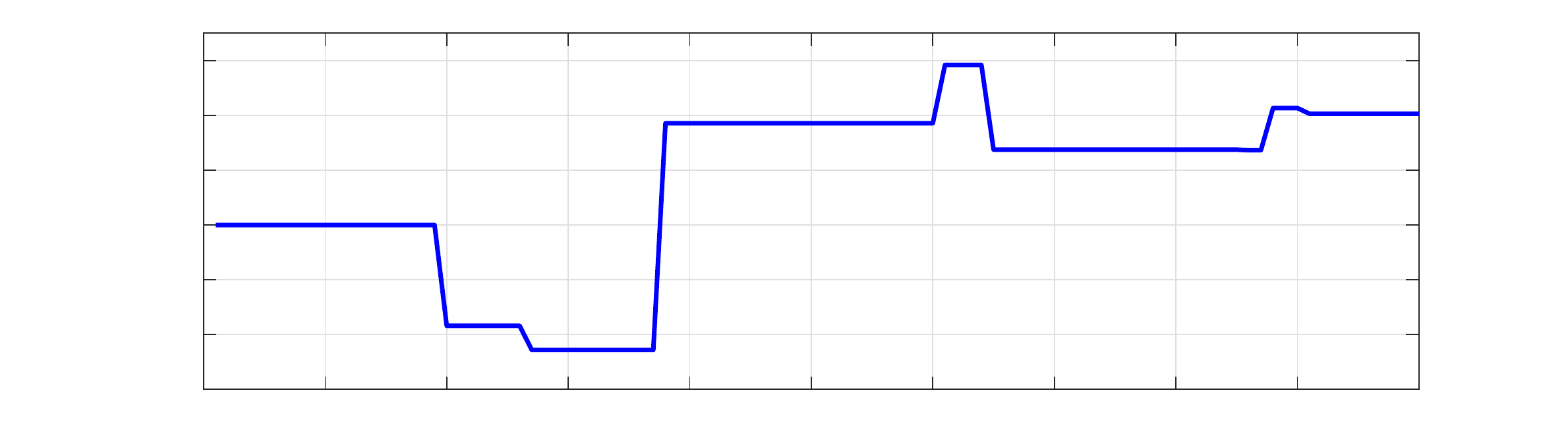}}

    \subfloat[\added{Student's t: $p(x) = \frac{\Gamma(\frac{\alpha+1}{2})}{\Gamma\big(\frac{\alpha}{2}\big)}\frac{1}{\sqrt{\pi}(1 + x^2)^{\frac{\alpha+1}{2}}}$}]{\includegraphics[trim={2.5cm 0 2cm 0}, clip,width=\columnwidth]{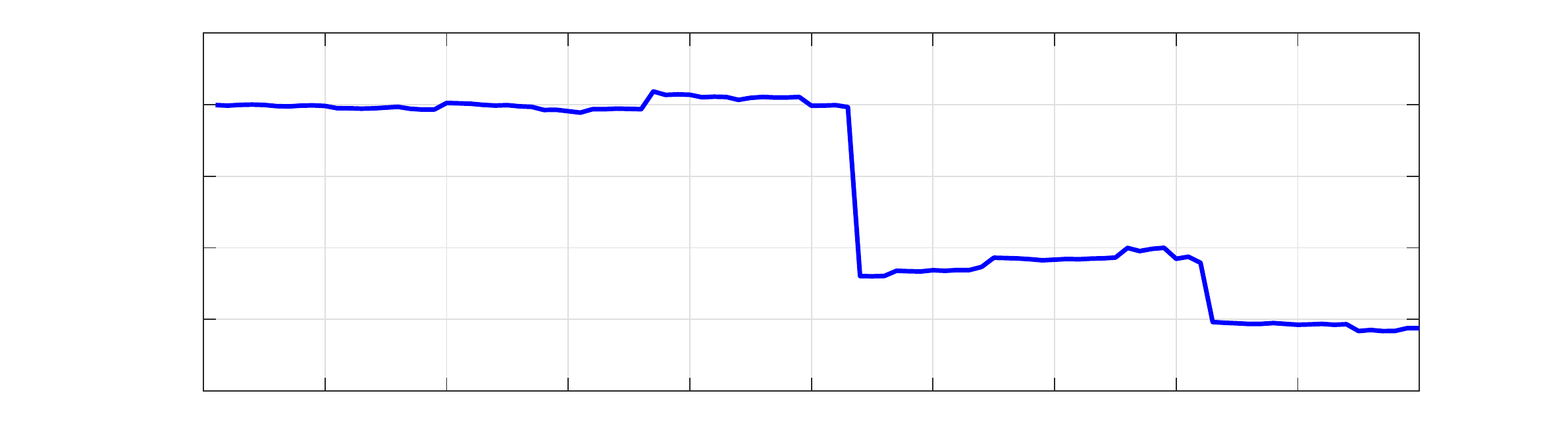}}

    \caption{Realizations of different L\'{e}vy processes as characterized by the corresponding infinitely divisible pdfs.}
    \label{fig:levy_processes}
\end{figure}

L\'{e}vy processes are closely linked to infinitely divisible (id) distributions. 

\begin{definition}[Infinite divisibility]
    A random variable $X$ is infinitely divisible if, for any $N \in \N\setminus\{0\}$, there exist independent and identically distributed (i.i.d.) random variables $X_1, \ldots, X_N$ such that $X = X_1 + \cdots + X_N$. 
\end{definition}

For any L\'{e}vy process $s$, the random variable $s(t)$ for some $t>0$ is infinitely divisible. Moreover, its probability density function (pdf) is given by
\begin{equation}
    p_{s(t)}(x) = \int_{\R} \bigg(\int_{\R} p_{s(1)}(y) \mathrm{e}^{\mathrm{j} \omega y} \ud y \bigg)^t \mathrm{e}^{-\mathrm{j} \omega x} \frac{\ud \omega}{2\pi}. 
\end{equation}  
Conversely, for any id distribution with pdf $p_{\text{id}}$, it is possible to construct a L\'{e}vy process $s$ such that $p_{s(1)} = p_{\text{id}}$. Thus, there is a one-to-one correspondence between L\'{e}vy processes and id distributions \cite{ken1999levy}. 

Among all id distributions, the pdf of the Gaussian distribution exhibits the fastest rate of decay at infinity. In this sense, we refer to the non-Gaussian, heavier-tailed members (\textit{e.g.,} Laplace, Bernoulli-Laplace, Student's t, symmetric-alpha-stable) of the class of id distributions as sparse \cite{amini2014sparsity}. Indeed, some of these sparse distributions have a mass at the origin in their probability distribution (\textit{e.g.,} Bernoulli-Laplace) and some of them are strongly compressible (\textit{e.g.,} Student's t, symmetric-alpha-stable) \cite{amini2011compressibility}. 

The stochastic model of L\'{e}vy processes allows us to consider a variety of signals with different types of sparsity. In our framework, we focus on the subclass of L\'{e}vy processes associated with the Gaussian, Laplace, Bernoulli-Laplace and Student's t distributions. Some realizations of these processes are shown in Figure \ref{fig:levy_processes}.

\subsection{Discrete Stochastic Model}
Now, we derive the pdf of the random vector $\M s = (s(k))_{k=1}^{K}$, which contains uniform samples of a L\'{e}vy process. Consider the stationary increment process $u(t) = \{s(t) - s(t-1) : t \in \R^{+} \}$ whose first-order pdf $p_u$ is the same as $p_{s(1)}$ and so is infinitely divisible. Its samples $\M u = (u(k))_{k=1}^{K}$ can be expressed as 
\begin{equation}\label{eq:discrete_innovation_model}
    \M u = \M D \M s,
\end{equation}
where $\M D$ is a finite-difference matrix of the form
\begin{equation}
    \M D = \begin{bmatrix}
        \phantom{-}1 & \phantom{-}0 & \phantom{-}0  & \phantom{-}\cdots  & \phantom{-}0\\
        -1 & \phantom{-}1 & \phantom{-}0  & \phantom{-}\cdots  & \phantom{-}0\\
        \phantom{-}0 & -1 & \phantom{-}1  & \phantom{-}\cdots  & \phantom{-}0\\
         &  & \phantom{-}\ddots & \phantom{-}\ddots &  \\
         \phantom{-}0 & \phantom{-}0  & \phantom{-}\cdots & -1 & \phantom{-}1
    \end{bmatrix}.
\end{equation}
Using \eqref{eq:discrete_innovation_model} and the fact that the increments are independent, we obtain the pdf of the discrete signal as
\begin{equation}\label{eq:discrete_pdf}
    p_{\M s}(\M s) = \prod_{k = 1}^{K} p_{u}\big([\M D \M s]_k\big).
\end{equation}

Note that \eqref{eq:discrete_innovation_model} can also be written as
\begin{equation}\label{eq:gen_levy}
    [\M s]_k = \sum_{n=1}^{k} [\M u]_n, \ \ \ \ k = 1, \ldots, K,
\end{equation}
which gives us a direct way to generate samples of L\'{e}vy processes. 

\subsection{Extensions}
In this work, we have considered inverse problems involving 1D signals that are modelled as realizations of L\'{e}vy processes with increments that follow the Gaussian, Laplace, Bernoulli-Laplace and Student's t distributions. Our framework can further be extended in a straightforward manner to include the more general signal model of continuous-domain first-order autoregressive processes \cite[Chapter 7]{unser2014ssp} driven by white noises associated with the aforementioned distributions. These AR(1) processes yield a discrete stochastic model that is similar to the one described in \eqref{eq:discrete_pdf}. There, the application of a suitable transformation matrix to the discrete signal vector, which contains equidistant samples of the process, decouples it and generates a random vector (called the innovation or generalized increments) with i.i.d. entries. Thus, the MMSE estimation methods presented in Section \ref{sec:mmse_gibbs} can be readily adapted for such AR(1) processes.

We can also directly extend the proposed framework to handle multidimensional signals for the particular stochastic model of continuous-domain AR L\'{e}vy sheets \cite[Chapter 3]{fageot2017gaussian}, \cite{unser2014ssp} associated with the Gaussian, Laplace, Bernoulli-Laplace and Student's t distributions. These are higher-dimensional generalizations (based on separable whitening operators) of the corresponding AR(1) processes and they result in desirable discrete models of the form \eqref{eq:discrete_pdf}. Unfortunately, the vectorized discrete signal for other (``non-separable") higher-dimensional stochastic processes described in \cite{unser2014ssp} cannot be fully decoupled by applying a linear transformation. This makes the task of designing schemes to compute their MMSE estimators very challenging. An alternate way of extending our framework could be to define a new class of continuous-domain multidimensional stochastic models using the spline-operator-based framework of \cite{unser2022ridges}. However, this approach would require substantial development of novel mathematical ideas and is thus not discussed further in this paper.

\section{Bayesian Inference}\label{sec:bayesian_inference}
So far, we have introduced the signal and measurement models that allow us to generate our ground-truth signals and simulate their noisy measurements for a certain acquisition setup. Next, we focus on statistical estimators for the reconstruction problem at hand, which is to recover the signal $\M s$ from the measurements $\M y$.

In Bayesian inference, the goal is to characterize the posterior distribution $p_{\M s|\M y}$ and derive estimators based on it. Using Bayes' rule and $\eqref{eq:discrete_pdf}$, we get
\begin{align}\label{eq:posterior}
    p_{\M s|\M y}(\M s | \M y) &= \frac{p_{\M y|\M s}(\M y | \M s) p_{\M s}(\M s)}{\int_{\R^K} p_{\M y|\M s}(\M y | \M s) p_{\M s}(\M s) \ud \M s}  \nonumber \\
    &\propto \exp\bigg(-\frac{\|\M y - \M H \M s\|_{2}^{2}}{2 \sigma_{\mathrm{n}}^{2}}\bigg) \prod_{k = 1}^{K} p_{u}\big([\M D \M s]_k\big).
\end{align}

\subsection{Maximum a Posteriori Estimator}
The MAP estimator calculates the mode of the posterior distribution $p_{\M s| \M y}$ and is given by 
\begin{align} \label{eq:MAP}
    \widehat{\M s}_{\text{MAP}}(\M y) &= \arg \max_{\M s \in \R^K} p_{\M s | \M y}(\M s | \M y) \nonumber \\
&= \arg \min_{\M s \in \R^K} \bigg(\frac{1}{2 \sigma_{\mathrm{n}}^{2}} \|\M y - \M H \M s\|_{2}^{2} + \sum_{k=1}^{K} \Phi_{u} \big([\M D \M s]_k \big) \bigg),
\end{align}
where $\Phi_u(x) = -\log(p_u(x))$. The cost functional in \eqref{eq:MAP} consists of a quadratic data-fidelity term and a penalty term that encodes the prior signal model. The optimization task in \eqref{eq:MAP} resembles the one formulated in the variational model-based methods. For instance, if $p_u$ is a Gaussian pdf, then the penalty term is proportional to $\|\M D \M s\|_{2}^{2}$, which is a classical Tikhonov regularizer \cite{tikhonov1963}. However, if $p_u$ is a Laplace pdf, then we have a sparsity-promoting $\ell_1$-norm penalty term $\|\M D \M s\|_{1}$, which corresponds to the popular TV regularizer \cite{rudin1992nonlinear}.

These MAP estimators can be computed efficiently with the help of iterative algorithms such as gradient descent, FISTA \cite{beck2009fast}, and ADMM \cite{boyd2011distributed}.

\subsection{Minimum Mean-Square Error Estimator}
The MMSE estimator is given by
\begin{align}\label{eq:MMSE}
    \widehat{\M s}_{\text{MMSE}}(\M y) &= \arg \min_{\widehat{\M s} \in \R^K} \bigg(\int_{\R^K} \|\M s - \widehat{\M s}\|_{2}^{2} \ p_{\M s|\M y}(\M s | \M y) \ud \M s \bigg) \nonumber \\
&= \int_{\R^K} \M s \ p_{\M s|\M y}(\M s | \M y) \ud \M s,
\end{align}
which is the mean of the posterior distribution $p_{\M s|\M y}$. For a fixed stochastic model, the MMSE estimator is the optimal reconstructor in the MSE sense and thus serves as the goldstandard in our benchmarking framework.

In the Gaussian case, the MMSE estimator is known to coincide with the MAP estimator and is straightforward to calculate \cite{kay1993fundamentals,unser2019biomedical}. However, in the non-Gaussian case, we need to numerically evaluate the high-dimensional integral in \eqref{eq:MMSE}, which is computationally challenging.

\section{MMSE Estimators for Sparse L\'{e}vy Processes}\label{sec:mmse_gibbs}
In this section, we present efficient methods to compute the MMSE estimator for sparse L\'{e}vy processes with increments that follow the Laplace, Student's t, and Bernoulli-Laplace distributions, which constitutes a key contribution of this paper.

\subsection{Markov Chain Monte Carlo Methods}
The MMSE estimator $\widehat{\M s}_{\text{MMSE}}$ involves the calculation of the integral \eqref{eq:MMSE}. The high dimensionality of this integral makes its approximation by simple techniques such as uniform-grid-based Riemann sums infeasible. Instead, one can use Markov Chain Monte Carlo (MCMC) methods \cite{hastings1970monte, geyer1992practical,gilks1995markov,gamerman2006markov} for the numerical approximation of \eqref{eq:MMSE} in a tractable manner.

MCMC methods are designed for generating random samples from nontrivial high-dimensional probability distributions. Broadly speaking, the idea in MCMC is to design a Markov chain such that the distribution that one wishes to draw samples from is its stationary distribution. The desired samples can be obtained by simulating the Markov chain and recording its states after convergence.

In order to compute the integral in \eqref{eq:MMSE}, we first generate samples $\{\M s^{(q)}\}_{q=1}^{Q}$ from $p_{\M s | \M y}$ using an MCMC method. We then approximate $\widehat{\M s}_{\text{MMSE}}$ by the empirical mean $\widehat{\M s}_{Q} = \frac{1}{Q} \sum_{q=1}^{Q} \M s^{(q)}$. Although the collected samples are correlated, the Markov chain central limit theorem \cite{gordin1978central} guarantees that $\widehat{\M s}_{Q}$ is a good approximation of $\widehat{\M s}_{\text{MMSE}}$ for a large-enough $Q$.

\subsection{Gibbs Sampling}
In this work, we propose to use the MCMC method called Gibbs sampling \cite{geman1984stochastic,casella1992explaining} to generate samples $\{\M u^{(q)}\}_{q=1}^{Q}$ from the posterior distribution $p_{\M u | \M y}$. These can then be transformed in accordance with \eqref{eq:gen_levy} to obtain samples $\{\M D^{-1} \M u^{(q)}\}_{q=1}^{Q}$ from $p_{\M s | \M y}$. We now give the gist of this algorithm. 

Let $x$ and $y$ be two random variables. Consider the task of generating samples from their joint distribution $p_{x, y}$. Gibbs sampling is advantageous whenever it is computationally difficult to sample from the joint distribution directly but the conditional distributions $p_{x | y}$ and $p_{y |x}$ are easy to sample from. The steps involved in this method are presented in Algorithm \ref{alg:gibbs_sampling}. They yield a Markov chain whose stationary distribution is indeed $p_{x, y}$ \cite{casella1992explaining}. In practice, one discards some of the initial samples (burn-in period) to allow the chain to converge. Moreover, quantities (expectation integrals) based on the marginal distributions $p_{x}$ and $p_{y}$ can be computed from the individual samples $\{x^{(q)}\}_{q=1}^{Q}$ and $\{y^{(q)}\}_{q=1}^{Q}$, respectively.  

\begin{algorithm}[t]
    \caption{Gibbs sampling}
    \begin{algorithmic}[1]
      \State \textbf{Input:} $Q$ (number of samples), $B$ (burn-in period)
      \State \textbf{Initialization:} $\big(\widetilde{x}^{(0)}, \widetilde{y}^{(0)}\big)$ 
      
      \For{$q = 1, \ldots, B+Q$}
        \State Generate $\widetilde{x}^{(q)} \sim p_{x | y}\big(x | \widetilde{y}^{(q-1)}\big)$
        \State Generate $\widetilde{y}^{(q)} \sim p_{y | x}\big(y | \widetilde{x}^{(q)}\big)$
      \EndFor

      \State \textbf{Output:} $\Big\{\big(x^{(q)}, y^{(q)}\big) \Big\}_{q=1}^{Q} = \Big\{\big( \widetilde{x}^{(q+B)}, \widetilde{y}^{(q+B)}  \big) \Big\}_{q=1}^{Q}$
    \end{algorithmic}
    \label{alg:gibbs_sampling}
  \end{algorithm}

Next, we present Gibbs sampling schemes for L\'{e}vy processes with Laplace, Student's t, and Bernoulli-Laplace increments. Our strategy is to introduce an auxiliary vector $\M w$ and perform Gibbs sampling for the joint distribution $p_{\M u, \M w | \M y}$ \cite{tanner1987calculation,mira1997use}. The key is to choose $\M w$ such that the conditional distributions $p_{\M u | \M w, \M y}$ and $p_{\M w | \M u, \M y}$ can be sampled from in an efficient manner. 

Hereafter, we assume that the noise variance $\sigma_{\mathrm{n}}^{2}$ and the parameters of the signal model are known.

\subsection{Laplace Increments}
For L\'{e}vy processes with Laplace increments, we adapt the approach that was developed in \cite{park2008bayesian}.  

The pdf for the Laplace distribution is 
\begin{equation}\label{eq:pdf_laplace}
    p_u(u) = \frac{b}{2}\exp\big(-b |u|\big),
\end{equation}
where $b$ is the scale parameter. The density in \eqref{eq:pdf_laplace} can be expressed as a scale mixture of normal distributions \cite{andrews1974scale}, as 
\begin{equation}
    p_u(u) = \int_{\R} p_{u|w}(u|w) p_w(w) \ud w,
\end{equation}
where
\begin{equation}\label{eq:normal_laplace}
    p_{u|w}(u|w) = \frac{1}{\sqrt{2 \pi w}} \exp\bigg(-\frac{u^2}{2 w}\bigg)
\end{equation}
is the Gaussian pdf and  
\begin{equation}\label{eq:mixing_density_laplace}
    p_w(w) = \frac{b^2}{2}\exp\bigg(-\frac{b^2 w}{2}\bigg) \mathbbm{1}_{+}(w)
\end{equation}
is a mixing exponential pdf\footnote{The pdf of the exponential distribution is \begin{equation*} p_{\text{exp}}(x) = (1/\lambda) \mathrm{e}^{-x/\lambda} \mathbbm{1}_{+}(x), \end{equation*} where $\lambda > 0$ is the scale parameter.} with $\lambda=2/b^2$. This property allows us to define an auxiliary random vector $\M w \in \R^K$ with i.i.d.\ entries following the distribution $p_w$ in \eqref{eq:mixing_density_laplace}, such that 
\begin{equation}\label{eq:aux_var_property_laplace}
    p_{\M u | \M w}(\M u | \M w) = \prod_{k=1}^{K} p_{u|w}\big([\M u]_k | [\M w]_k\big),
\end{equation}
where $p_{u|w}$ is shown in \eqref{eq:normal_laplace}.

Due to the chain rule of probability (or the general product rule), the full joint distribution $p_{\M y, \M u, \M w}$ can be written as 
\begin{align}\label{eq:full_joint_distribution_laplace}
    p_{\M y, \M u, \M w}(\M y, \M u, \M w) &=  p_{\M y | \M u, \M w}(\M y | \M u, \M w)  p_{\M u, \M w}(\M u, \M w) \nonumber \\
    &= p_{\M y | \M u}(\M y | \M u) p_{\M u | \M w}(\M u | \M w) p_{\M w}(\M w).
\end{align}
Consequently, the distribution $p_{\M u, \M w | \M y}$ takes the form
\begin{align}\label{eq:conditional_joint_distribution_laplace}
    p_{\M u, \M w | \M y}(\M u, \M w | \M y) \propto \ &\exp\bigg(-\frac{1}{2 \sigma_{\mathrm{n}}^2} \|\M y - \M A \M u \|_{2}^{2}\bigg) \nonumber \\
    &\times \ \prod_{k=1}^{K}[\M w]_k^{-\frac{1}{2}} \exp\bigg(-\frac{[\M u]_k^2}{2 [\M w]_k}\bigg) \nonumber \\
    &\times \ \prod_{k=1}^{K} \frac{b^2}{2}\exp\bigg(-\frac{b^2 [\M w]_k}{2}\bigg) \mathbbm{1}_{+}\big([\M w]_k\big),
\end{align}
where $\M A \coloneqq \M H \M D^{-1}$. 

Based on \eqref{eq:conditional_joint_distribution_laplace}, the conditional distribution $p_{\M u | \M w, \M y}$ is then obtained as
\begin{align}\label{eq:conditional_u_laplace}
    p_{\M u | \M w, \M y}(\M u | \M w, \M y) \propto \ \exp\bigg(-\frac{1}{2}\bigg(&\frac{1}{\sigma_{\mathrm{n}}^2}\|\M y - \M A \M u\|_{2}^{2} \nonumber \\
    & + \M u^{T} \M C_{\mathrm{L}}(\M w) \M u \bigg) \bigg),
\end{align}
where $\M C_{\mathrm{L}}(\M w)$ is a diagonal matrix with elements $\big([\M w]_k^{-1}\big)_{k=1}^{K}$. Specifically, $p_{\M u | \M w, \M y}$ is a multivariate Gaussian pdf with mean $\overline{\M u} = \sigma_{\mathrm{n}}^{-2} \big(\sigma_{\mathrm{n}}^{-2} \M A^{T} \M A + \M C_{\mathrm{L}}(\M w)\big)^{-1}\M A^{T}\M y$ and covariance matrix $\overline{\M R} = \big(\sigma_{\mathrm{n}}^{-2} \M A^{T} \M A + \M C_{\mathrm{L}}(\M w)\big)^{-1}$. There exist several methods for the efficient generation of samples from a multivariate Gaussian density \cite{rue2001fast, orieux2012sampling, gilavert2014efficient, vono2022high}.

The conditional distribution $p_{\M w | \M u, \M y}$ is
\begin{equation}
    p_{\M w | \M u, \M y}(\M w | \M u, \M y) \propto \prod_{k=1}^{K} p_{w|u, \M y}\big([\M w]_k|[\M u]_k, \M y\big),
\end{equation}
where
\begin{align}\label{eq:conditional_w_laplace}
    p_{w | u, \M y}\big(w | u, \M y\big) \propto \ &\exp\bigg(-\frac{1}{2}\bigg(\frac{u^2}{w} + b^2 w\bigg)\bigg) \nonumber \\
    &\times \ w^{-\frac{1}{2}} \mathbbm{1}_{+}(w)
\end{align}
belongs to the family of generalized inverse Gaussian distributions\footnote{The pdf of the generalized inverse Gaussian distribution is \begin{equation*} p_{\text{gig}}(x) = \frac{(\lambda_1/\lambda_2)^{a/2}}{2K_a(\sqrt{\lambda_1 \lambda_2})}x^{a-1} \mathrm{e}^{-(\lambda_1 x + \lambda_2/x)/2} \mathbbm{1}_{+}(x), \end{equation*} where $K_a$ is the modified Bessel function of the second kind, $\lambda_1 > 0$, $\lambda_2 > 0$, and $a \in \R$.} with $\lambda_1 = b^2$, $\lambda_2 = u^2$ and $a = 0.5$. We use the method proposed in \cite{devroye2014random} to draw samples from the pdf in \eqref{eq:conditional_w_laplace}.

To summarize, at each iteration $q$ of the constructed blocked Gibbs sampler, we generate $\M u^{(q)} \sim p_{\M u | \M w,\M y}\big(\M u | \M w^{(q-1)}, \M y\big)$ and $[\M w^{(q)}]_k \sim p_{w | u,\M y}\big(w | [\M u^{(q)}]_k,\M y \big)$ for all $k\in \{1,\ldots,K\}$. The collected samples $\{\M u^{(q)}\}_q$ follow the desired distribution $p_{\M u | \M y}$.

\subsection{Student's t Increments}
The case of Student's t increments can be handled by adapting the method shown in \cite{fevotte2006bayesian}, which is in fact similar to the one we described for Laplace increments.

The Student's t pdf is given by 
\begin{equation}\label{eq:pdf_student}
    p_u(u) = \frac{\Gamma(\frac{\alpha+1}{2})}{\Gamma\big(\frac{\alpha}{2}\big)}\frac{1}{\sqrt{\pi}(1 + u^2)^{\frac{\alpha+1}{2}}},
\end{equation}
where $\alpha$ is the number of degrees of freedom and controls the tail of the distribution, and where $\Gamma$ denotes the gamma function. It can also be expressed as 
\begin{equation}
    p_u(u) = \int_{\R} p_{u|w}(u|w) p_w(w) \ud w,
\end{equation}
where
\begin{equation}\label{eq:normal_student}
    p_{u|w}(u|w) = \sqrt{\frac{w}{2 \pi}} \exp\bigg(- \frac{w u^2}{2}\bigg)
\end{equation}
is a Gaussian pdf and  
\begin{equation}\label{eq:mixing_density_student}
    p_w(w) = \frac{(0.5)^{\frac{\alpha}{2}}}{\Gamma(\frac{\alpha}{2})}w^{\frac{\alpha}{2}-1} \exp{\Big(-\frac{w}{2}\Big)} \ \mathbbm{1}_{+}(w)
\end{equation}
is the pdf of a gamma\footnote{The pdf of the gamma distribution is \begin{equation*} p_{\text{gam}}(x) = \frac{1}{\lambda_2^{\lambda_1} \Gamma(\lambda_1)}x^{\lambda_1-1}\mathrm{e}^{-x/\lambda_2} \mathbbm{1}_{+}(x), \end{equation*} where $\lambda_1 > 0$ and $\lambda_2 > 0$ are the shape and scale parameters, respectively.} distribution. Again, we introduce an auxiliary vector $\M w \in \R^K$ whose i.i.d.\ entries follow $p_w$ defined in \eqref{eq:mixing_density_student}. It is such that 
\begin{equation}\label{eq:aux_var_property_student}
    p_{\M u | \M w}(\M u | \M w) = \prod_{k=1}^{K} p_{u|w}\big([\M u]_k | [\M w]_k\big),
\end{equation}
where $p_{u|w}$ is defined in \eqref{eq:normal_student}.

Here, the distribution $p_{\M u, \M w | \M y}$ is given by
\begin{align}\label{eq:conditional_joint_distribution_student}
    p_{\M u, \M w | \M y}(\M u, \M w | \M y) \propto \ &\exp\bigg(-\frac{1}{2 \sigma_{\mathrm{n}}^2} \|\M y - \M A \M u \|_{2}^{2}\bigg) \nonumber \\
    &\times \ \prod_{k=1}^{K}[\M w]_k^{\frac{1}{2}} \exp\bigg(-\frac{[\M w]_k [\M u]_k^2}{2}\bigg) \nonumber \\
    &\times \ \prod_{k=1}^{K} [\M w]_k^{\frac{\alpha}{2} - 1} \exp{\bigg(-\frac{[\M w]_k}{2}\bigg)} \mathbbm{1}_{+}\big([\M w]_k\big),
\end{align}
where $\M A \coloneqq \M H \M D^{-1}$. 

Now, the conditional distribution $p_{\M u | \M w, \M y}(\M u | \M w, \M y)$ turns out to be
\begin{align}\label{eq:conditional_u_student}
    p_{\M u | \M w, \M y}(\M u | \M w, \M y) \propto \ \exp\bigg(-\frac{1}{2}\bigg(&\frac{1}{\sigma_{\mathrm{n}}^2}\|\M y - \M A \M u\|_{2}^{2} \nonumber \\
    & + \M u^{T} \M C_{\mathrm{T}}(\M w) \M u \bigg) \bigg),
\end{align}
where $\M C_{\mathrm{T}}(\M w)$ is a diagonal matrix with entries $\big([\M w]_k\big)_{k=1}^{K}$. Similar to the Laplace case, $p_{\M u | \M w, \M y}$ is a multivariate Gaussian density with mean $\overline{\M u} = \sigma_{\mathrm{n}}^{-2} \big(\sigma_{\mathrm{n}}^{-2} \M A^{T} \M A + \M C_{\mathrm{T}}(\M w)\big)^{-1}\M A^{T}\M y$ and covariance matrix $\overline{\M R} = \big(\sigma_{\mathrm{n}}^{-2} \M A^{T} \M A + \M C_{\mathrm{T}}(\M w)\big)^{-1}$. 

The distribution $p_{\M w | \M u, \M y}$ is again separable and takes the form 
\begin{equation}
    p_{\M w | \M u, \M y}(\M w | \M u, \M y) \propto \prod_{k=1}^{K} p_{w|u, \M y}\big([\M w]_k|[\M u]_k, \M y\big),
\end{equation}
where
\begin{align}\label{eq:conditional_w_student}
    p_{w | u, \M y}\big(w | u, \M y\big) \propto \ &\exp\bigg(-\frac{(1 + u)^2 w}{2}\bigg) \nonumber \\
    &\times \ w^{\frac{\alpha-1}{2}} \mathbbm{1}_{+}(w).
\end{align}
is a gamma distribution with $\lambda_1 = \frac{\alpha + 1}{2}$ and $\lambda_2 = \frac{2}{(1 + u)^2}$, which can easily be sampled from.

\subsection{Bernoulli-Laplace Increments}
In \cite{ge2011enhanced}, Gibbs sampling schemes have been designed for a deconvolution problem where the underlying signal is an i.i.d.\ spike train that follows the Bernoulli-Gaussian distribution. Unfortunately, the Bernoulli-Gaussian distribution is not infinitely divisible and so is not compatible with our framework of L\'{e}vy processes. While there exists some work \cite{chaari2013sparse} on Bernoulli-Laplace priors, according to the analysis presented in \cite{ge2011enhanced}, their proposed sampler would have a tendency to get stuck in certain configurations. Thus, we build upon the method in \cite{ge2011enhanced} and develop a novel Gibbs sampler for L\'{e}vy processes with Bernoulli-Laplace increments.

The Bernoulli-Laplace pdf is
\begin{equation}\label{eq:pdf_bl}
    p_u(u) = \lambda \delta(u) + (1-\lambda)\frac{b}{2}\exp{\big(-b |u|\big)},  
\end{equation}
where $\lambda \in (0,1)$ denotes the mass probability at the origin and $b$ is a scale parameter. We can represent this same density as
\begin{equation}
    p_u(u) = \int_{\R} \bigg(\sum_{v=0}^{1} p_{u|v, w}(u|v, w) p(v) \bigg) p(w) \ud w,
\end{equation}
where
\begin{equation}\label{eq:bernoulli_bl}
    p_{v}(v) = (\lambda)^{1-v}(1-\lambda)^{v} \ \ \text{for } v \in \{0,1\}
\end{equation}
is a Bernoulli distribution,
\begin{equation}\label{eq:bernoulli_exponential}
    p_{w}(w) = \frac{b^2}{2}\exp\bigg(-\frac{b^2 w}{2}\bigg) \mathbbm{1}_{+}(w)
\end{equation}
is an exponential pdf, and $p_{u|v, w}$ is defined such that
\begin{align}
    &p_{u|v, w}(u|v=0, w) = \delta(u) \label{eq:dirac_bl}\\
    &p_{u|v, w}(u|v=1, w) = \frac{1}{\sqrt{2 \pi w}} \exp\bigg(-\frac{u^2}{2 w}\bigg) \label{eq:gaussian_bl}.
\end{align}
Based on this representation, we introduce two independent auxiliary vectors $\M v \in \R^K$ and $\M w \in \R^K$. Their elements are i.i.d.\ and follow the distributions $p_{v}$ and $p_{w}$, as defined in \eqref{eq:bernoulli_bl} and \eqref{eq:bernoulli_exponential}, respectively. Further, these vectors satisfy
\begin{equation}\label{eq:aux_var_property_bl}
    p_{\M u | \M v, \M w}(\M u | \M v, \M w) = \prod_{k=1}^{K}p_{u|v, w}\big([\M u]_k | [\M v]_k, [\M w]_k\big),
\end{equation}
where $p_{u|v, w}$ is defined in \eqref{eq:dirac_bl} and \eqref{eq:gaussian_bl}.

Here, the full joint distribution $p_{\M y, \M u, \M v, \M w}$ is given by 
\begin{align}
    p_{\M y, \M u, \M v, \M w}(\M y, \M u, \M v, \M w) = \ &p_{\M y | \M u, \M v, \M w}(\M y | \M u, \M v, \M w) p_{\M u, \M v, \M w}(\M u, \M v, \M w) \nonumber \\
    = \ &p_{\M y | \M u}(\M y | \M u) p_{\M u | \M v, \M w}(\M u | \M v, \M w) \nonumber \\ 
    &\times \ p_{\M v}(\M v) p_{\M w}(\M w).
\end{align}
As a result, the distribution $p_{\M u, \M v, \M w|\M y}$ takes the form
\begin{align}\label{eq:conditional_joint_distribution_bl}
    p_{\M u, \M v, \M w | \M y}(\M u, \M v, \M w | \M y&) \ \propto \ \exp\bigg(-\frac{1}{2 \sigma_{\mathrm{n}}^2} \|\M y - \M A \M u \|_{2}^{2}\bigg) \nonumber \\
    &\times \ \prod_{k=1}^{K}p_{u|v, w}\big([\M u]_k | [\M v]_k, [\M w]_k\big) \nonumber \\
    &\times \ \prod_{k=1}^{K} \lambda^{1-[\M v]_k}(1-\lambda)^{[\M v]_k} \nonumber \\
    &\times \ \prod_{k=1}^{K} \frac{b^2}{2}\exp\bigg(-\frac{b^2 [\M w]_k}{2}\bigg) \mathbbm{1}_{+}([\M w]_k),
\end{align}
where $\M A = \M H \M D^{-1}$. 

Let us now introduce some notations. For any binary vector $\M q \in \R^K$, let $\mathcal{I}_{\M q, 0}$ and $\mathcal{I}_{\M q, 1}$ denote sets of indices such that $[\M q]_k = 0$ for $k \in \mathcal{I}_{\M q, 0}$ and $[\M q]_k = 1$ for $k \in \mathcal{I}_{\M q, 1}$. Further, let $\M A(\M q)$ be the matrix constructed by taking the columns of $\M A$ corresponding to the indices in $\mathcal{I}_{\M q, 1}$. We then define the matrix $\M B(\M q, \M r) = \sigma_{\mathrm{n}}^{2} \M I + \M A(\M q) \M C_{\mathrm{BL}}(\M q, \M r) \M A(\M q)^{T}$, where $\M r \in \R^K$ is a vector with positive entries and $\M C_{\mathrm{BL}}(\M q, \M r)$ is a diagonal matrix with entries $([\M r]_k)_{k \in \mathcal{I}_{\M q, 1}}$. Here, we also introduce the vector $\M q_{(-k)} \in \R^{K-1}$ that contains all the entries of $\M q$ except the $k$th one, so that $\M q_{(-k)} = ([\M q]_{1}, \ldots, [\M q]_{k-1}, [\M q]_{k+1}, \ldots, [\M q]_{K})^T$. Lastly, for $q \in \{0,1\}$, we define the vector $\M q_{(-k)}^{q} \in \R^{K}$ such that $\M q_{(-k)}^{q} = ([\M q]_{1}, \ldots, [\M q]_{k-1}, q, [\M q]_{k+1}, \ldots, [\M q]_{K})^T$. 

First, we look at the conditional distribution $p_{\M u | \M v, \M w, \M y}$. From \eqref{eq:dirac_bl} and \eqref{eq:conditional_joint_distribution_bl}, we deduce that any sample from $p_{\M u | \M v, \M w, \M y}$ takes the value of zero at the indices in $\mathcal{I}_{\M v, 0}$. If we define $\M u_1 = ([\M u]_k)_{k \in \mathcal{I}_{\M v, 1}}$, then we get
\begin{align}\label{eq:conditional_u_bl}
    p_{\M u_1 | \M v, \M w, \M y}(\M u_1 | \M v, \M w, \M y) \propto  \exp\bigg(\hspace{-0.1cm}&-\frac{1}{2}\bigg(\frac{1}{\sigma_{\mathrm{n}}^2}\|\M y - \M A(\M v) \M u_1\|_{2}^{2} \nonumber \\
    & + \M u_1^{T} \M C_{\mathrm{BL}}(\M v, \M w) \M u_1 \bigg) \bigg).
\end{align}
Thus, $p_{\M u_1 | \M v, \M w, \M y}$ is a multivariate Gaussian density with mean $\overline{\M u_1} = \sigma_{\mathrm{n}}^{-2} \big(\sigma_{\mathrm{n}}^{-2} \M A(\M v)^{T} \M A(\M v) + \M C_{\mathrm{BL}}(\M v, \M w)\big)^{-1}\M A(\M v)^{T}\M y$ and covariance matrix $\overline{\M R} = \big(\sigma_{\mathrm{n}}^{-2} \M A(\M v)^{T} \M A(\M v) + \M C_{\mathrm{BL}}(\M v, \M w)\big)^{-1}$.

The conditional distribution $p_{\M w | \M u, \M v, \M y}$ takes the form
\begin{align}
    p_{\M w | \M u, \M v, \M y}(\M w | \M u, \M v, &\M y) \propto \prod_{k=1}^{K} p_{w|u, v, \M y}\big([\M w]_k|[\M u]_k, [\M v]_k, \M y \big), 
\end{align} 
where $p_{w|u, v, \M y}$ is given by
\begin{equation}\label{eq:conditional_w2_bl_exp}
    p_{w|u, v, \M y}(w|u, v=0, \M y) \propto \frac{b^2}{2}\exp\bigg(-\frac{b^2 w}{2}\bigg) \mathbbm{1}_{+}(w)
\end{equation}
\begin{align}\label{eq:conditional_w2_bl_gig}
    p_{w|u, v, \M y}(w|u, v=1, \M y) \propto \ &\exp\bigg(-\frac{1}{2}\bigg(\frac{u^2}{w} + b^2 w\bigg)\bigg) \nonumber \\
    &\times \ w^{-\frac{1}{2}} \mathbbm{1}_{+}(w).
\end{align}
The densities in \eqref{eq:conditional_w2_bl_exp} and \eqref{eq:conditional_w2_bl_gig} correspond to the exponential distribution with $\lambda = 2/b^2$ and the generalized inverse Gaussian distribution with $\lambda_1 = b^2$, $\lambda_2 = u^2$, and $a = 0.5$.

Next, inspired by the work in \cite{ge2011enhanced}, we consider sampling from the marginalized conditional distribution of $[\M v]_k$ in a sequential manner as this can allow for a more efficient exploration of configurations of $\M v$. More specifically, at each iteration $q$, we draw $[\M v^{(q)}]_k$ from the distribution $p_{[\M v]_k | \M v_{(-k)}, \M w, \M y}\big(v | \M v_{(-k)}^{(q)}, \M w^{(q)}, \M u^{(q-1)}\big)$, where $\M v_{(-k)}^{(q)} = \big([\M v^{(q)}]_{1}, \ldots, [\M v^{(q)}]_{k-1}, [\M v^{(q-1)}]_{k+1}, \ldots, [\M v^{(q-1)}]_{K} \big)$ and $k \in \{1, \ldots, K\}$.

The marginalized posterior distribution $p_{\M v, \M w | \M y}$ is given by 
\begin{align}\label{eq:marginal_integral_bl_tmp}
    p_{\M v, \M w | \M y}(\M v, \M w | \M y) \propto p_{\M y|\M v, \M w}(\M y|\M v, \M w) p_{\M v}(\M v) p_{\M w}(\M w), 
\end{align}
where 
\begin{equation}\label{eq:marginal_integral_bl}
    p_{\M y|\M v, \M w}(\M y|\M v, \M w) = \hspace{-0.15cm} \int_{\R^K} p_{\M y|\M u, \M v, \M w}(\M y | \M u, \M v, \M w) p_{\M u|\M v, \M w}(\M u|\M v, \M w) \ud \M u.
\end{equation}
It can be shown that \eqref{eq:marginal_integral_bl_tmp} and \eqref{eq:marginal_integral_bl} lead to
\begin{align}\label{eq:marginalized_posterior_distribution_bl}
    p_{\M v, \M w | \M y}(\M v, \M w | \M y) \propto& \ |\M B(\M v, \M w)|^{-\frac{1}{2}} \exp\bigg(-\frac{1}{2} \M y^{T} \M B(\M v, \M w)^{-1} \M y \bigg) \nonumber \\
    &\times \ \prod_{k=1}^{K} \lambda^{1-[\M v]_k}(1-\lambda)^{[\M v]_k} \nonumber \\
    &\times \ \prod_{k=1}^{K} \frac{b^2}{2}\exp\bigg(-\frac{b^2 [\M w]_k}{2}\bigg) \mathbbm{1}_{+}([\M w]_k).
\end{align}
From \eqref{eq:marginalized_posterior_distribution_bl}, we see that $p_{[\M v]_k | \M v_{(-k)}, \M w, \M y}$ is a Bernoulli distribution with
\begin{align*}
    p_{[\M v]_k | \M v_{(-k)}, \M w, \M y}(v|\M v_{(-k)}, \M w, \M y) = \hspace{4cm}
\end{align*}
\begin{align} 
    \bigg(1 + \exp\bigg(-\frac{1}{2}\Big(&h\big(1-v; \M v_{(-k)}, \M w, \M y\big) \nonumber \\
    &- h\big(v; \M v_{(-k)}, \M w, \M y\big)\Big)\bigg)\bigg)^{-1},
\end{align}
where
\begin{align}
    h\big(v; \M v_{(-k)}, \M w, \M y\big) =& \ \M y^{T} \M B\big(\M v_{(-k)}^{v}, \M w\big)^{-1} \M y \nonumber \\
    &+ \log\big(|\M B\big(\M v_{(-k)}^{v}, \M w\big)|\big) \nonumber \\
    &+ 2 v \log\Big(\frac{\lambda}{1-\lambda}\Big).
\end{align}

To summarize, in each iteration $q$ of the above-described sampler, we generate $\M w^{(q)} \sim p_{\M w | \M u, \M v, \M y}\big(\M w | \M u^{(q-1)}, \M v^{(q-1)}, \M y\big)$, $[\M v^{(q)}]_k \sim p_{[\M v]_k | \M v_{(-k)}, \M w, \M y}\big(v | \M v_{(-k)}^{(q)}, \M w^{(q)}, \M u^{(q-1)}\big)$ for all $k$ and $\M u^{(q)} \sim p_{\M u | \M v, \M w,\M y}\big(\M u | \M v^{(q)}, \M w^{(q)}, \M y\big)$. This particular order of updates is important as it yields a partially collapsed Gibbs sampler \cite{van2008partially} where the stationary distribution is still $p_{\M u, \M v, \M w | \M y}$.

\section{Experimental Results}\label{sec:experiments}
In our experiments, we benchmark the performance of some popular signal reconstruction schemes, including a CNN-based method, on deconvolution and Fourier sampling problems with L\'{e}vy processes associated with the Bernoulli-Laplace and Student's t distributions.

\subsection{Signal Models}
We consider a signal vector $\M s \in \R^{100}$ that contains samples of a L\'{e}vy process whose increments follow the Bernoulli-Laplace or Student's t distribution.

\subsubsection{Bernoulli-Laplace increments}
The Bernoulli-Laplace pdf \eqref{eq:pdf_bl} is characterized by the parameters $\lambda$ and $b$, where $\lambda$ determines the mass probability at the origin and $b$ represents the scale of the Laplace component. We perform experiments for models corresponding to $\lambda \in \{ 0.6, 0.7, 0.8, 0.9 \}$. The scale parameter is set to $b=1$ for each case.

\subsubsection{Student't t increments}
The Student's t pdf \eqref{eq:pdf_student} is parameterized by $\alpha$, which controls the tails of the distribution. We conduct experiments for $\alpha \in \{1, 3, 5, 39 \}$.

\subsection{Measurement Models}
We consider both deconvolution and Fourier sampling problems for each of the above-described signal models.

\subsubsection{Deconvolution}
As shown in Section \ref{sec:deconv_forward}, the system matrix $\M H$ for deconvolution is a discrete convolution matrix. Accordingly, we construct $\M H: \R^{100} \rightarrow \R^{88}$ such that
\begin{equation}
    \M H = \begin{bmatrix}
        [\M h]_{13} & \cdots & [\M h]_1 & 0  & \cdots & 0 \\
        0 & \ddots & & \ddots & \ddots & \vdots \\ 
        \vdots & \ddots & \ddots & & \ddots & 0 \\
        0 & \cdots & 0 & [\M h]_{13} & \cdots & [\M h]_1
        \end{bmatrix},
\end{equation}
where $\M h \in \R^{13}$ consists of the central samples of a truncated Gaussian PSF with variance $\sigma_{0}^{2} = 4$.

\subsubsection{Fourier Sampling}
\textcolor{black}{For Fourier sampling in 1D, which is reminiscent of MRI, the forward model $\M H$ resembles a discrete Fourier matrix (see Section \ref{sec:fourier_sampling_forward}). Thus, in order to construct $\M H$, we first sample $M' = 16$ rows of the DFT matrix.} The first row of the DFT matrix (DC component) is always kept, while the remaining ones are selected in a quasi-random fashion with a denser sampling at low frequencies. We then create the real system matrix $\M H: \R^{100} \rightarrow \R^{M}$, where $M = 2M' - 1$, by separating the real and imaginary parts.

In both measurement models, the AWGN variance $\sigma_{\mathrm{n}}^{2}$ is chosen such that the (average) measurement SNR is around $30$ dB.

\subsection{Reconstruction Algorithms}
For each combination of the signal and measurement models, we compare the performance of some (variational) model-based techniques, a CNN-based scheme and the MMSE estimator. We generate validation and test datasets, each consisting of $1,\!000$ pairs of ground-truth signals and their noisy measurements. Further, in order to train the CNNs, we also synthesize a repository $\mathcal{R}$ containing a large number of training examples. 

{\color{black}
\subsubsection{Model-based methods}
We consider the model-based methods 
\begin{equation}\label{eq:est_l2}
    \widehat{\M s}_{\ell_2} = \arg \min_{\M s \in \R^K} \Big(\|\M y - \M H \M s\|_{2}^{2} + \tau \|\M D \M s\|_{2}^{2} \Big),
\end{equation}
\begin{equation}\label{eq:est_l1}
    \widehat{\M s}_{\ell_1} = \arg \min_{\M s \in \R^K} \Big(\|\M y - \M H \M s\|_{2}^{2} + \tau \|\M D \M s\|_{1} \Big),
\end{equation}
and
\begin{equation}\label{eq:est_log}
    \widehat{\M s}_{\text{log}} = \arg \min_{\M s \in \R^K} \Big(\|\M y - \M H \M s\|_{2}^{2} + \tau \sum_{k=1}^{K} \log \Big(1 + \big([\M D \M s]_k\big)^{2}\Big) \Big),
\end{equation}
where $\tau \in \R_{+}$. Equations \eqref{eq:est_l2}, \eqref{eq:est_l1} and \eqref{eq:est_log} resemble the MAP estimators of L\'{e}vy processes associated with Gaussian, Laplace, and Student's t distributions, respectively. However, unlike the MAP estimators, these include an adjustable hyperparameter $\tau$. For each of these methods, the same regularization parameter $\tau$ is used for the entire test dataset. This particular value of $\tau$ is the one that yields the lowest MSE for the validation dataset. 

These estimators are implemented in MATLAB using GlobalBioIm \cite{soubies2019pocket}---a library for solving inverse problems. Specifically, the $\ell_2$ estimator is expressed in closed-form as
\begin{equation}
    \widehat{\M s}_{\ell_2} = \big(\M H^{T} \M H + \tau \M D^{T} \M D \big)^{-1} \M H^{T} \M y.
\end{equation}
The $\ell_1$ and log estimators are computed iteratively using ADMM. Since the cost functional in \eqref{eq:est_log} is non-convex, we initialize ADMM for $\widehat{\M s}_{\text{log}}$ with $\widehat{\M s}_{\ell_1}$ so that it can reach a better local minimum. 
}

\begin{figure}[t]
    \centering
    \includegraphics[width=\linewidth]{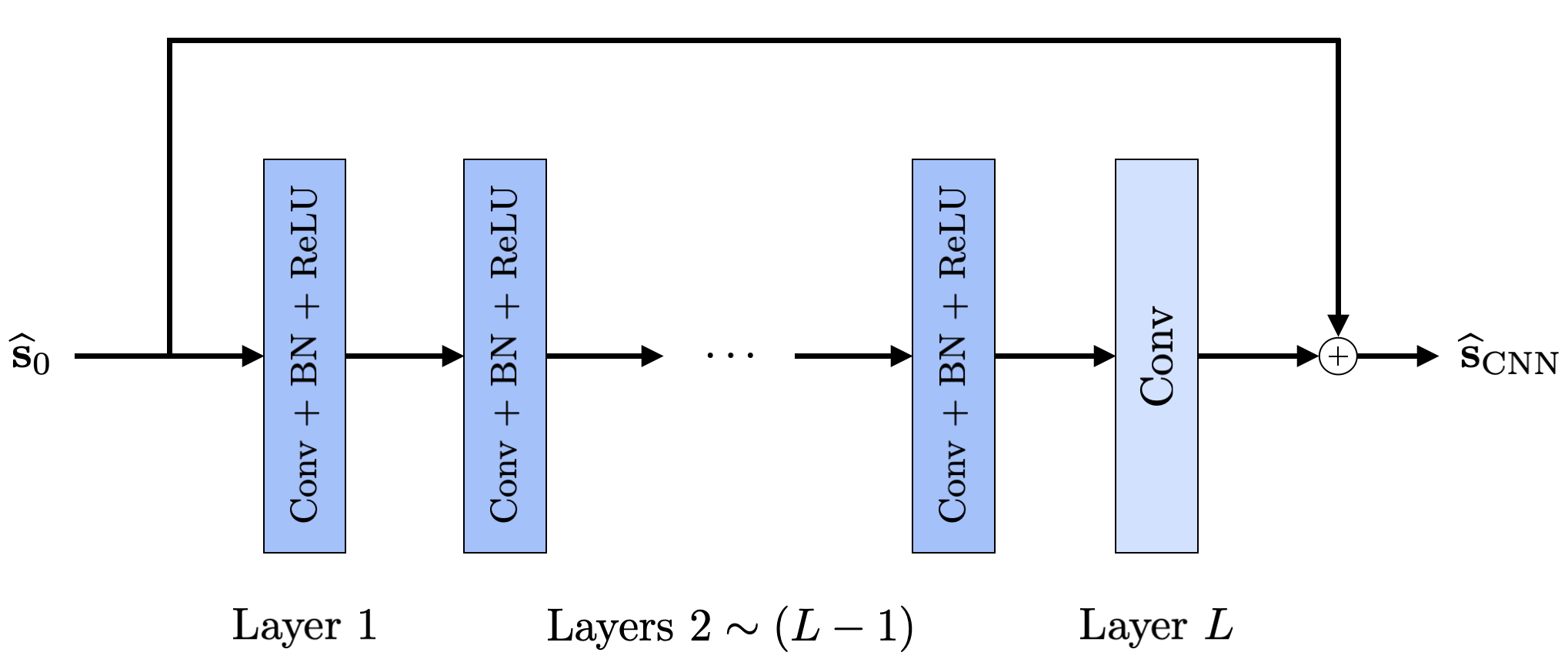}
    \caption{Architecture of the CNN, where BN denotes the operation of batch normalization.}
    \label{fig:CNN_architecture}
\end{figure}
        
\begin{table}[t]
\caption{Convolution Layers.}
\label{table:conv_layers}
\setlength{\tabcolsep}{3pt}
\renewcommand{\arraystretch}{1.5}
\centering
\begin{tabular}[c]{c| c | c | c}
\hline \hline
Layer  & Filter size & {Input channels} & Output channels  \\ \hline 
1 & $(F \times 1)$ & $1$ & $C$ \\
2 $\sim$ ($L-1$) & $(F \times 1)$ & $C$ & $C$  \\ 
$L$ & $(F \times 1)$ & $C$ & $1$  \\ \hline \hline 
\end{tabular}
\end{table}

\subsubsection{CNN-based method}
The concept here is to train a CNN as a regressor that maps an initial low-quality reconstruction $\widehat{\M s}_{0}$ to a high-quality one $\widehat{\M s}_{\text{CNN}}$ \cite{jin2017deep,chen2017low,hyun2018deep, monakhova2019learned,perdios2021cnn}. The architecture of the CNN used in our experiments is based on the well-known denoising network DnCNN \cite{zhang2017beyond} and is described in Figure \ref{fig:CNN_architecture} and Table~\ref{table:conv_layers}. 

First, we build a training dataset of cardinality $M_T$ by taking the first $M_T$ examples $\{\M s_{m}, \M y_m\}_{m=1}^{M_T}$ from the repository $\mathcal{R}$. We then train the model by minimizing the MSE loss function 
\begin{equation}
    \mathcal{L}(\V \theta) = \frac{1}{M_T} \sum_{m=1}^{M_T} \|\M s_{m} - \widehat{\M s}_{\text{CNN}}\big(\V \theta; \widehat{\M s}_{0}(\M y_m)\big) \|_{2}^{2},
\end{equation}
where $\V \theta$ represents the learnable parameters of the network, with the help of the ADAM optimizer \cite{kingma2014adam}. The CNN is trained for $1,\!000$ epochs with a batch size of $256$ and a weight decay of $\gamma$. The initial learning rate is set as $10^{-2}$. For some duration of the training (first $600$ epochs for deconvolution and first $750$ epochs for Fourier sampling), it is decreased by a factor of $0.5$ every $50$ epochs. We choose the initial low-quality reconstruction to be $\widehat{\M s}_{0}(\M y) = \M H^{T} \M y$ for the deconvolution problems. In the case of Fourier sampling, $\widehat{\M s}_{0}(\M y)$ is the zero-filled reconstruction. All the CNN-based reconstruction schemes are implemented in PyTorch.

\subsubsection{Goldstandard (MMSE estimator)}
Our MMSE estimators are implemented in MATLAB, according to the methods detailed in Section \ref{sec:mmse_gibbs}. There, we set the number of samples as $Q=8,\!000$ and the burn-in period as $B=3,\!000$ for signals with Bernoulli-Laplace increments. We use $Q=15,\!000$ and $B=5,\!000$ for signals associated with the Student's t distribution. 

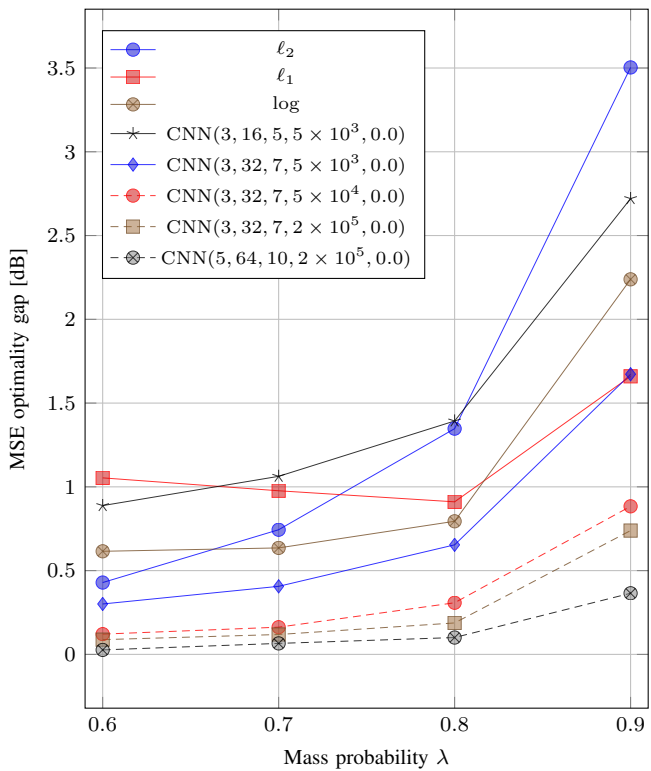
\begin{figure}[t]
    \centering
    \begin{tikzpicture}
    \begin{axis}[
        width = 0.5\textwidth,
        height = 0.6\textwidth,
        xlabel = {Mass probability $\lambda$},
        ylabel = {MSE optimality gap~[dB]},
        grid = both,
        grid style = {thin},
        xtick = {0.6,0.7,0.8,0.9},
        legend pos = north west,
        legend style={font=\scriptsize,},
        xmin=0.59, xmax=0.91,
	]
	\foreach \column in {l2,l1,log,NN1, NN2, NN3, NN4, NN5}{
        \addplot +[fill opacity=0.5, draw opacity=0.8, line width=0.05pt, mark size=2.5pt] table
[x=lam,y expr=\thisrow{\column}-\thisrow{MMSE}]{figures/rev_deconv_bl_data.txt};

    }
    \legend{$\ell_2$,$\ell_1$,$\log$,{$\mathrm{CNN}(3,16,5,5 \times 10^{3},0.0)$}, {$\mathrm{CNN}(3,32,7,5 \times 10^{3},0.0)$}, {$\mathrm{CNN}(3,32,7,5 \times 10^{4},0.0)$}, {$\mathrm{CNN}(3,32,7,2 \times 10^{5},0.0)$}, {$\mathrm{CNN}(5,64,10,2 \times 10^{5},0.0)$}};
    \end{axis}
\end{tikzpicture}
    \caption{Deconvolution of L\'{e}vy processes with Bernoulli-Laplace increments.}
    \label{fig:deconvolution_bl}
\end{figure}

\begin{figure}[t]
    \centering
    \begin{tikzpicture}
    \begin{axis}[
        width = 0.5\textwidth,
        height = 0.6\textwidth,
        xlabel = {Mass probability $\lambda$},
        ylabel = {MSE optimality gap~[dB]},
        grid = both,
        grid style = {thin},
        xtick = {0.6,0.7,0.8,0.9},
        legend pos = north west,
        legend style={font=\scriptsize,},
        xmin=0.59, xmax=0.91,
	]
	\foreach \column in {l2,l1,log,NN1, NN2, NN3}{
        \addplot +[fill opacity=0.5, draw opacity=0.8, line width=0.05pt, mark size=2.5pt] table
[x=lam,y expr=\thisrow{\column}-\thisrow{MMSE}]{figures/rev_fourier_sampling_bl_data.txt};
    }
    \legend{$\ell_2$,$\ell_1$,$\log$,{$\mathrm{CNN}(5,64,10,2 \times 10^{5}, 5 \times 10^{-3})$}, {$\mathrm{CNN}(9,64,15,5 \times 10^{5}, 5 \times 10^{-4})$}, {$\mathrm{CNN}(9,96,20,7.5 \times 10^{5}, 5 \times 10^{-4})$}};
    \end{axis}
\end{tikzpicture}
    \caption{Fourier sampling of L\'{e}vy processes with Bernoulli-Laplace increments.}
    \label{fig:fourier_sampling_bl}
\end{figure}
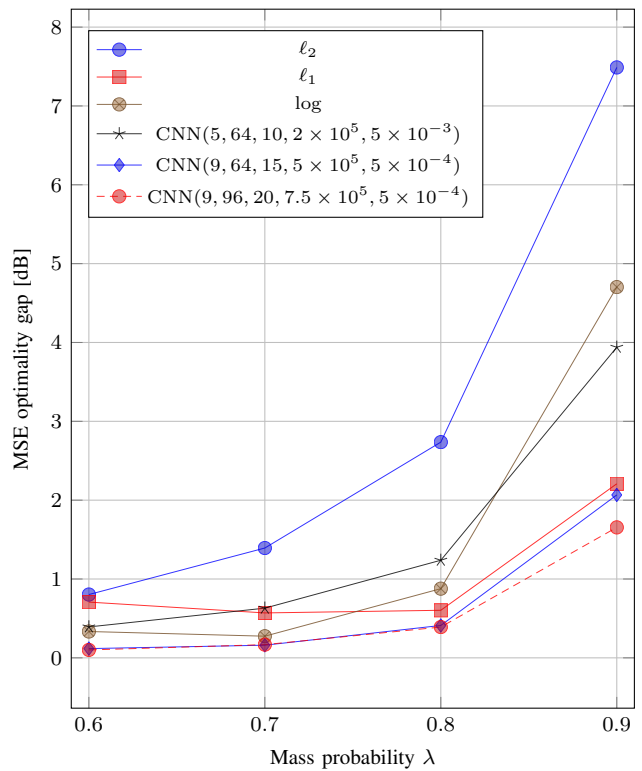

\begin{figure}[!t]
    \centering
    \begin{tikzpicture}
    \begin{semilogxaxis}[
        width = 0.5\textwidth,
        height = 0.6\textwidth,
        xlabel = {Degree of freedom $\alpha$},
        ylabel = {MSE optimality gap~[dB]},
        grid = both,
        grid style = {thin},
        xtick = {39, 5, 3, 1},
        legend style={font=\scriptsize,at={(0.025,0.975)},anchor=north west},
  		x dir = reverse,
  		x tick label style={log ticks with fixed point}]
	\foreach \column in {l2, l2_star, l1, l1_star, log, log_star, NN2, NN3, NN5}{
            \addplot +[fill opacity=0.5, draw opacity=0.8, line width=0.05pt, mark size=2.5pt] table[x=alph,y expr=\thisrow{\column}-\thisrow{MMSE}]{figures/rev_deconv_student_data.txt};
        }

        \draw[dashed,line width=1.25pt ] (axis cs:2.5,-1) rectangle (axis cs:45,2.5);
    
    \legend{$\ell_2$,$\ell_2^{*}$,$\ell_1$,$\ell_1^{*}$,$\log$,$\log^{*}$,{$\mathrm{CNN}(3,32,7,5 \times 10^{3},0.0)$}, {$\mathrm{CNN}(3,32,7,5 \times 10^{4},0.0)$}, {$\mathrm{CNN}(5,64,10,2 \times 10^{5},0.0)$}}
    \end{semilogxaxis}
\end{tikzpicture}

\vspace{0.4cm}

\begin{tikzpicture}
    \begin{semilogxaxis}[
        width = 0.5\textwidth,
        height = 0.5\textwidth,
        xlabel = {Degree of freedom $\alpha$},
        ylabel = {MSE optimality gap~[dB]},
        grid = both,
        grid style = {thin},
        xtick = {39, 5, 3},
        legend style={font=\scriptsize,at={(0.025,0.975)},anchor=north west},
  		x dir = reverse,
  		x tick label style={log ticks with fixed point},
  		xmin=2.5, xmax= 45]
	\foreach \column in {l2, l2_star, l1, l1_star, log, log_star, NN2, NN3, NN5}{
            \addplot +[fill opacity=0.5, draw opacity=0.8,restrict x to domain=1.08:3.67, line width=0.05pt, mark size=2.5pt] table[x=alph,y expr=\thisrow{\column}-\thisrow{MMSE}]{figures/rev_deconv_student_data.txt};
        }
          \end{semilogxaxis}


\end{tikzpicture}

    \caption{Deconvolution of L\'{e}vy processes with Student's t increments. The figure at the bottom is a zoomed-in version of the dotted rectangular box shown in the figure at the top.}
    \label{fig:deconvolution_student}
\end{figure}

\begin{figure}[!t]
    \centering
    \begin{tikzpicture}
    \begin{semilogxaxis}[
        width = 0.5\textwidth,
        height = 0.6\textwidth,
        xlabel = {Degree of freedom $\alpha$},
        ylabel = {MSE optimality gap~[dB]},
        grid = both,
        grid style = {thin},
        xtick = {39, 5, 3, 1},
        legend style={font=\scriptsize,at={(0.025,0.975)},anchor=north west},
  		x dir = reverse,
  		x tick label style={log ticks with fixed point},
	]
	\foreach \column in {l2, l2_star, l1, l1_star, log, log_star, NN1, NN2, NN3}{
        \addplot +[fill opacity=0.5, draw opacity=0.8, line width=0.05pt, mark size=2.5pt]
table[x=alph,y expr=\thisrow{\column}-\thisrow{MMSE}]{figures/rev_fourier_sampling_student_data.txt};
    }
    
    \draw[dashed,line width=1.25pt ] (axis cs:2.5,-1) rectangle (axis cs:45,2.5);
    
    \legend{$\ell_2$,$\ell_2^{*}$,$\ell_1$,$\ell_1^{*}$,$\log$,$\log^{*}$,{$\mathrm{CNN}(5,64,10,2 \times 10^{5},10^{-3})$}, {$\mathrm{CNN}(7,64,12, 2 \times 10^{5},5 \times 10^{-3})$}, {$\mathrm{CNN}(9,64,15,3.5 \times 10^{5},10^{-3})$}};
    \end{semilogxaxis}

\end{tikzpicture}

\vspace{0.4cm}

\begin{tikzpicture}
    \begin{semilogxaxis}[
        width = 0.5\textwidth,
        height = 0.5\textwidth,
        xlabel = {Degree of freedom $\alpha$},
        ylabel = {MSE optimality gap~[dB]},
        grid = both,
        grid style = {thin},
        xtick = {39, 5, 3, 1},
        legend style={font=\scriptsize,at={(0.025,0.975)},anchor=north west},
  		x dir = reverse,
  		x tick label style={log ticks with fixed point},
  		xmin=2.5, xmax= 45,
	]
	\foreach \column in {l2, l2_star, l1, l1_star, log, log_star, NN1, NN2, NN3}{
        \addplot +[fill opacity=0.5, draw opacity=0.8,restrict x to domain=1.08:3.67, line width=0.05pt, mark size=2.5pt]
table[x=alph,y expr=\thisrow{\column}-\thisrow{MMSE}]{figures/rev_fourier_sampling_student_data.txt};
    }
    \end{semilogxaxis}

\end{tikzpicture}

    \caption{Fourier sampling of L\'{e}vy processes with Student's t increments. The figure at the bottom is a zoomed-in version of the dotted rectangular box shown in the figure at the top.}
    \label{fig:fourier_sampling_student}
\end{figure}

\subsection{Results}
We present our results for all the test datasets in Figures \ref{fig:deconvolution_bl}, \ref{fig:fourier_sampling_bl}, \ref{fig:deconvolution_student} and \ref{fig:fourier_sampling_student}. For the sake of clarity, instead of the MSE, we display the ``MSE optimality gap" which is the difference between the MSE obtained by a specific method and the MSE attained by the MMSE estimator. In these figures, the CNNs are labelled using the tuple $(F, C, L, M_T, \gamma)$, where $F$ is the filter size, $C$ is the number of channels, $L$ is the number of layers, $M_T$ is the cardinality of the training dataset and $\gamma$ is the weight decay.

\subsubsection{L\'{e}vy processes with Bernoulli-Laplace increments}
Here, we summarize our observations for both the deconvolution and Fourier sampling experiments (Figures \ref{fig:deconvolution_bl} and \ref{fig:fourier_sampling_bl}).

The sparsity-promoting $\ell_1$ estimator, which corresponds to the popular TV regularization, is known to be well-suited to piecewise-constant L\'{e}vy processes with Bernoulli-Laplace increments. As the value of $\lambda$ increases, these signals become sparser and exhibit fewer jumps. Consequently, we observe that the $\ell_1$ estimator performs better than the $\ell_2$ estimator. The log estimator also promotes sparse solutions \cite{wipf2010iterative} and we see that it performs well for these piecewise-constant signals. However, despite the good fit, there is still some gap between the MSE attained by the $\ell_1$ and log, and MMSE estimators.

The performance of the CNN-based method improves as we increase the capacity of the CNN and the amount of training data. In fact, with sufficient capacity and training data, they outperform the $\ell_1$ and log estimators. Remarkably, some of the CNNs achieve a near-optimal MSE.

\subsubsection{L\'{e}vy processes with Student's t increments}
The parameter $\alpha$ allows us to consider a wide range of signals. As $\alpha \rightarrow \infty$, we approach the Gaussian regime. The other extreme is $\alpha=1$, which corresponds to the super heavy-tailed (sparse) Cauchy distribution. This scenario can be particularly challenging for the correct setting of algorithm parameters. Due to the heavy tails of the Cauchy distribution, the validation and test datasets may contain signals with a vastly different range of values. Consequently, for a given model-based method, the regularization parameter $\tau$ that is chosen to yield the lowest MSE for the validation dataset may differ significantly from the value $\tau^{*}$ that achieves the lowest MSE on the test dataset. Thus, in Figures \ref{fig:deconvolution_student} and \ref{fig:fourier_sampling_student}, we also include the performance of model-based methods when their regularization parameter is tuned for optimal MSE performance on the test dataset directly. These ``boosted" model-based methods are labelled as $\ell_2^{*}$, $\ell_1^{*}$ and log$^{*}$.

In Figures \ref{fig:deconvolution_student} and \ref{fig:fourier_sampling_student}, we can see that the $\ell_2$ estimator is optimal for a large value of $\alpha$. As the value of $\alpha$ decreases, the performance of the $\ell_2$ estimator deteriorates and becomes worse than that of the $\ell_1$ estimator. For all the cases, the log estimator, which corresponds to a tunable MAP estimator for the Student's t distribution, attains reasonable MSE values. Note that for the deconvolution experiment involving Cauchy signals, there is a significant gap between the MSE values obtained by the $\ell_2$ and $\ell_1$ and $\ell_2^{*}$ and $\ell_1^{*}$ estimators, respectively. Interestingly, the log estimator is less affected by this issue.

Finally, for both deconvolution and Fourier sampling problems, CNNs with sufficient capacity and training data perform well up to $\alpha = 3$, after which there seems to be a steep transition and their performance drops sharply. In fact, for Cauchy signals, we observe that the training process for these CNNs is quite unstable---the training loss marginally decreases and seems to converge, and the networks do not generalize to the validation (or test) datasets. We believe that this last example poses an open challenge for designing robust neural-network-based schemes that can handle signals following (super) heavy-tailed distributions.

\section{Conclusion}
We have introduced a controlled environment, based on sparse stochastic processes (SSPs), for the objective benchmarking of reconstruction algorithms, including CNN-based methods that require lots of training data, in the context of linear inverse problems. We have developed efficient posterior sampling schemes to compute the minimum-mean-square-error estimators for specific classes of SSPs. These yield the upper limit on reconstruction performance and allow us to provide a measure of statistical optimality. We have highlighted the abilities of our framework by benchmarking some popular variational methods and convolutional neural-network (CNN) architectures for deconvolution and Fourier-sampling problems. In particular, we have observed that, while CNNs outperform the variational methods and achieve a near-optimal performance in terms of mean-square error for a wide range of conditions, they can sometimes fail too, especially for signals with heavy-tailed innovations.

\section*{Acknowledgements}
We acknowledge access to the facilities and expertise of the CIBM Center for Biomedical Imaging, a Swiss research center of excellence founded and supported by Lausanne University Hospital (CHUV), University of Lausanne (UNIL), École polytechnique fédérale de Lausanne (EPFL), University of Geneva (UNIGE), and Geneva University Hospitals (HUG).

\bibliographystyle{IEEEtran.bst} 
\bibliography{refs}

\end{document}